\documentclass [aps,prb,preprint] {revtex4}
\usepackage{latexsym, epsfig, graphics}
\begin{document}

\preprint{LBNL-63374}

\title{Examples of the Zeroth Theorem of the History of Science}
\author{J. D Jackson\thanks{Electronic  mail: jdjackson@lbl.gov}}
\affiliation{Physics Department, University of California, Berkeley \\ and Lawrence Berkeley National Laboratory, Berkeley, California 94720}
\date{10 October 2007}

\begin{abstract}
%(text of abstract goes here)
The zeroth theorem of the history of science (enunciated by E. P. Fischer) and widely known in the mathematics community as Arnol'd's Principle (decreed by M. V. Berry), states that a discovery (rule,regularity, insight) named after someone (often) did not originate with that person. I present five examples from physics: the Lorentz condition $\partial_{\mu}A^{\mu} = 0$ defining the Lorentz gauge of the electromagnetic potentials; the Dirac delta function $\delta (x)$; the Schumann resonances of the earth-ionosphere cavity; the Weizs\"{a}cker-Williams method of virtual quanta; the BMT equation of spin dynamics. I give illustrated thumbnail sketches of both the true and reputed discoverers and quote from their "discovery" publications.
\end{abstract}
\normalsize
\pacs{????}
\maketitle

\section*{ I. \ \ \ INTRODUCTION} 
% (Begin text of article)
\indent In a column entitled ``Fremde Federn. Im Gegenteil,'' (very loosely, ``Inappropriate Attributions'') in the July 24, 2006 issue of \emph{Die Welt}, a Berlin newspaper, historian of science Ernst Peter Fischer~\cite{fischer}  gave a name to a phenomenon of which some of us are aware, that sometimes (often?) a physical discovery or law or a number is attributed to and named after a person who is arguably not the first person to make the discovery:
\begin{quote}
``Das Nullte Theorem der Wissenschaftsgeschichte lauten, dass eine Entdeckung (Regel, Gesetzm\"{a}ssigkeit, Einsicht), die nach einer Person benannt ist, nicht von dieser Person herr\"{u}hrt.''~\cite{fischer}\\
(The zeroth theorem of the history of science reads that a discovery (rule, regularity, insight), named after someone, (\emph{often?}) did not originate with that person.) [\emph{often?} added]\\
\end{quote} 
Fischer goes on to give examples, some of which are \\
\indent Avogadro's \emph{number} ($6.022 \times 10^{23}$) was first determined by Loschmidt in 1865 (although Avogadro had found in 1811 that any gas at NTP had the same number of molecules per unit volume).\\
\indent Halley's comet was known 100 years before Halley noted its appearance at regular intervals and predicted correctly its next appearance).\\
\indent Olber's paradox (1826) was discussed by Kelper (1610) and by Halley and Cheseaux in the 18th century.\\
\indent Fischer spoke of examples in the natural sciences, as do I. But numerous instances exist in other areas. In mathematics the theorem is known as Arnol'd's principle or Arnol'd's law, after the Russian mathematician V. I. Arnol'd.  Arnol'd's law was enunciated by M. V. Berry~\cite{berry, arnold} some years ago to codify Arnol'd's efforts to correct inappropriate attributions that neglect Russian mathematicians. Berry also proposed Berry's law, a far broader, self-referential theorem: ``Nothing is ever discovered for the first time.''~\cite{berry}  If the Zeroth Theorem were known as Fischer's law, it would be a clear example of Arnol'd's principle.  As it is, the Zeroth Theorem stands as an illustration of Berry's law.\\
  
In each example I present the bare bones of the issue - the named effect, the generally recognized ``owner,'' the prior ``claimant'', with dates. After briefly describing the protagonists' origins and careers, I quote from the appropriate literature to establish the truth of the specific example.\\

\section*{II. \ \ \  THE LORENTZ CONDITION AND LORENTZ GAUGE FOR THE ELECTROMAGNETIC POTENTIALS} 
\indent My first example is the Lorentz condition that defines the Lorentz gauge for the electromagnetic potentials  $\varphi$ and ${\bf a}$.  The relation was specified by the Dutch theoretical physicist Hendrik Antoon Lorentz in 1904 in an encyclopedia article.~\cite{lorentz1}  In his notation the constraint reads:\\
\begin{equation}
div \ {\bf a} = - \frac{1}{c} \dot{\varphi} \label{lorentz}
\end{equation}
or, in covariant form,
\begin{equation}
\partial_{\mu}A^{\mu}\ =\ 0 \label{lorentzc}
\end{equation}
where $A^{\mu} = (\varphi,{\bf a})$ and $\partial_{\mu} = (\frac{\partial}{c \partial t}, {\bf \nabla})$.  Eq.(\ref{lorentz}) or (\ref{lorentzc}) is so famous and familiar that any citation of it will be to some textbook. If it is ever actually traced back to Lorentz, the reference will likely the cited encyclopedia  article or his book, \emph{Theory of Electrons},~\cite{lorentz2} published in 1909. 

Lorentz was not the first to point out Eq.(\ref{lorentz}). Thirty-seven years earlier, in 1867, the Danish theorist Ludvig Valentin Lorenz, writing about the identity of light with the electromagnetism of charges and currents,~\cite{lorenz}  stated the constraint on his choice of potentials.  His version of Eq.(\ref{lorentz}) reads:
\begin{equation}
\frac{d\bar{\Omega}}{d t} = - 2 \left(\frac{d\alpha}{d x} + \frac{d\beta}{d y} + \frac{d\gamma}{d z} \right) \label{lorenz}
\end{equation}
where $\bar{\Omega}$ is the scalar potential and $(\alpha, \beta, \gamma)$ are the components of the vector potential. The strange factors of 2 and 4 appearing here and below have their origins in a since abandoned definition of the electric current in terms of moving charges.

In 1900, in a Festschrift volume marking the 25th anniversary of H. A. Lorentz's doctorate, the Prussian theorist Emil Johann Wiechert described the introduction of the scalar and vector potentials into the Maxwell equations in much the way it is done in textbooks today.~\cite{wiechert}  He notes that the divergence of the vector potential is not constrained by the relation ${\bf B} = {\bf \nabla \times A} $ and imposes the condition, in his notation where the vector potential is ${\bf \Gamma}$ and $V$ is the speed of light:
\begin{equation}
\frac{\partial \phi}{\partial t}\ + \ V \left(\frac{\partial \Gamma_{x}}{\partial x} + \frac{\partial \Gamma_{y}}{\partial y} + 
\frac{\partial \Gamma_{z}}{\partial z} \right) \ = \ 0 \label{wiechert}
\end{equation}

\subsection*{II.A  \ \  \   Ludvig Valentin Lorenz (1829-1891)}
Ludvig Valentin Lorenz was born in 1829 in Helsing\o r, Denmark of German-Huguenot extraction. After gymnasium, in 1846 he entered the Polytechnic High School in Copenhagen, which had been founded by \O rsted in the year of Lorenz's birth.
He graduated as a chemical engineer from the University of Copenhagen in 1852.
With occasional teaching jobs, Lorenz pursued research in physics and in 1858 went to Paris to work with Lam\'{e} among others. An examination essay on elastic waves resulted in a paper of 1861, where retarded solutions of the wave equation were first published. On his return to Copenhagen, he published on optics (1863) and the identity of light with electromagnetism already mentioned. In 1866 he was appointed to the faculty of the Military High School outside Copenhagen and also elected a member of the Royal Danish Academy of Sciences and Letters.  After 21 years at the Military High School, Lorenz obtained the support of the Carlsberg Foundation from 1887 until his death in 1891. In 1890 his last paper (only in Danish) was a detailed treatment on the scattering of radiation by spheres, anticipating what is known as ``Mie scattering'' (1908), another example of the zeroth theorem, not included here.

\subsection*{II.B \ \ \  Emil Johann Wiechert (1861-1928) }
Emil Johann Wiechert was born in Tilsit, Prussia. When he was 18 he and his widowed mother moved to K\"{o}nigsberg where he attended first a gymnasium and then the Albertus university.  He completed his Ph.D. on elastic waves in 1889 and began as a lecturer and researcher in K\"{o}nigsberg in 1890. His research, both experimental and theoretical, encompassed electromagnetism, cathode rays, and electron theory, as well as geophysical topics such as the mass distribution within the earth. In 1897 he was invited to G\"{o}ttingen, first as \emph{Prizatdozent} and then in 1898 made Professor and Director of the G\"{o}ttingen Geophysical Institute, the first of its kind. Wiechert remained at G\"{o}ttingen for the rest of his career, training students and working in geophysics and seismology, with occasional forays into theoretical physics topic such as relativity, and electron theory. He found his colleagues, Felix Klein, David Hilbert, and his former mentor Woldemar Voigt, congenial and stimulating enough to turn down numerous offers of professorships elsewhere. In physics Wiechert's name is famous for the Li\'{e}nard-Wiechert potentials of a relativistic charged particle; in geophysics, for the Wiechert inverted-pendulum seismograph.

\subsection*{II.C  \ \ \  Hendrik Antoon Lorentz (1853-1928)}
Hendrik Antoon Lorentz was born in Arnhem, The Netherlands, in 1853. After high school in Arnhem, 1866-69, he attended the University of Leiden where he studied physics and mathematics, graduating in 1872. He received his Ph.D. in 1875 for a thesis on aspects on the electromagnetic theory of light. His academic career began in 1878 when at the age of 24 Lorentz was appointed Professor of Theoretical Physics at Leiden, a post he held for 34 years. His research ranged widely, but centered on electromagnetism and light. His name is associated with Lorenz in the Lorenz-Lorentz relation between the index of refraction of a medium and its density and composition (Lorenz, 1875; Lorentz, 1878). Notable were his works in the 1890s on the electron theory of electromagnetism (now called the microscopic theory, with charged particles at rest and in motion as the sources of the fields) and the beginnings of relativity (the FitzGerald-Lorentz length contraction hypothesis,  1895) and a bit later the Lorentz transformation. Lorentz shared the 1902 Nobel Prize in Physics with Pieter Zeeman for ``their researches into the influence of magnetism upon radiation phenomena.'' He received many honors and memberships in learned academies and was prominent in national and international scientific organizations until his death in 1928.\\
\begin{figure} [h]
\centering
\includegraphics[width=5in]{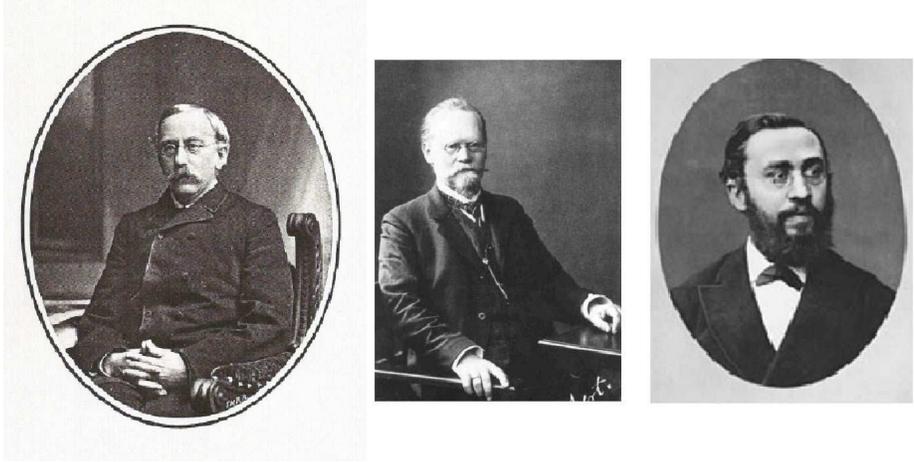}
\caption{(L-R) Ludvig V. Lorenz, Emil J. Wiechert, Hendrik A. Lorentz}
\end{figure}

\subsection*{II.D  \ \ \  Textual Evidence}
Lorenz's 1867 paper~\cite{lorenz} establishing the identity of light with electromagnetism was evidently written without knowledge of Maxwell's famous work of 1865. He begins with the quasi-static potentials, with the vector potential in the Kirchhoff-Weber form,~\cite{jackson-okun} and proceeds toward the differential equations for the fields. Actually, following the continental approach of Helmholtz, Lorenz uses electric current density instead of electric field, noting the connection via Ohm's law and the conductivity (called \emph{k}). 

But before proceeding to his full theory, Lorenz establishes that his generalization of the quasi-static limit is consistent with all known observations to date. Using a retarded form of the scalar potential, he demonstrates by expanding the retarded time $t^{\prime} = t-r/a $ in powers of $r/a$, where $a$ is a velocity parameter, that his retarded scalar potential and an emergent retarded vector potential yield the same electric field as the instantaneous Kirchhoff-Weber forms to second order in the presumably small $r/a$. Furthermore, by clever choices of the velocity $a$, he is able to show a restricted class of what are now called gauge transformations involving the Neumann and Kirchhoff-Weber forms of the vector potential, although he does not emphasize the point.
 
Because he is including light within his framework, he is not content with the quasi-static approximation. He proceeds to define the current density (electric field) components as $(u, v, w)$, write a retarded form for the scalar potential, called $\bar{\Omega}$, and then present the (almost) familiar expressions for the current density/electric field in terms of the scalar and vector potential: 
\begin{quote}
``Hence the equations for the propagation of electricity, as regards the experiments on which they rest, are just as valid as [the quasi-static equations] if [...] the following form be assigned to them,~\cite{partials}
\begin{eqnarray*}
u & = & -2k\left(\frac{d\bar{\Omega}}{dx}+\frac{4}{c^{2}}\frac{d\alpha}{dt} \right), \\
v & = & -2k\left(\frac{d\bar{\Omega}}{dy}+\frac{4}{c^{2}}\frac{d\beta}{dt} \right), \\
w & = & -2k\left(\frac{d\bar{\Omega}}{dz}+\frac{4}{c^{2}}\frac{d\gamma}{dt} \right), \\
\end{eqnarray*}
where, for brevity's sake, we put
\begin{eqnarray*}
\alpha & = & \int \int \int \frac{dx^{\prime} dy^{\prime} dz^{\prime}}{r}u^{\prime}(t-r/a) \\
\beta  & = & \int \int \int \frac{dx^{\prime} dy^{\prime} dz^{\prime}}{r}v^{\prime}(t-r/a) \\
\gamma & = & \int \int \int \frac{dx^{\prime} dy^{\prime} dz^{\prime}}{r}w^{\prime}(t-r/a) \\
\end{eqnarray*}
These equations are distinguished from equations (I) [the Kirchhoff-Weber forms] by containing, instead of $ U, V, W $, the somewhat less complicated members $\alpha, \beta, \gamma $; and they express further that the entire action between free electricity and the electric currents \emph{requires time to propagate itself} ........'' \\
\end{quote}
I judge from the throw-away phrases, ``are just as valid'' and ``for brevity's sake,'' and earlier remarks distinguishing his form of the vector potential from the Kirchhoff-Weber form (in addition to having retardation), that Lorenz understood gauge invariance without formally introducing the concept. In passing it is curious to note than in a paper published a year later~\cite{maxwell}, Maxwell criticized Lorenz's (and Riemann's) use of retarded potentials, claiming that they violated conservation of energy and momentum. But Lorenz had referred to his 1861 paper on the propagation of elastic waves to observe that the wave equation is satisfied by retarded sources.

In his march toward the differential equations for the ``fields,'' Lorenz notes that with his choice of the scalar and vector potentials:
\begin{quote}
``\ldots\ we obtain \\
%\begin{equation}
\[ \frac{d\bar{\Omega}}{d t} = - 2 \left(\frac{d\alpha}{d x} + \frac{d\beta}{d y} + \frac{d\gamma}{d z} \right) \ \]
%\end{equation}
\indent ``Moreover from (5),
%\begin{equation}
\[ \frac{1}{a^{2}} \frac{d^{2} \alpha}{d t^{2}} = \Delta_{2} \  \alpha + 4 \pi u \  \]
%\end{equation}
``and in like manner for $\beta, \gamma$.  \ldots''
\end{quote}
Lorenz then proceeds to derive the Amp\`{e}re-Maxwell equation relating the curl of the magnetic field to the sum of the displacement current and the conduction current density and goes on to obtain the other equations equivalent to Maxwell's.

In the last part of his paper, Lorenz sets himself the task of reversing his path, beginning with his differential equations for the fields, which he views as describing light, and working back toward his form of the retarded potentials. Imposition of Eq.(\ref{lorenz}) leads to the simple wave equations whose solutions, as he proved in 1861, are the standard retarded potentials in what is known as the Loren(t)z gauge. Lorenz stresses that:
\begin{quote}
``This result is a new proof of the identity of the vibrations of light with electrical currents; for it is clear now, not only that the laws of light can be deduced from those of electrical currents, but that the converse way may be pursued . . . .''
\end{quote}

In his 1900 Lorentz Festschrift paper,``Elektrodynamische  Elementargestze,''~\cite{wiechert} Wiechert  begins by summarizing the theory of optics, introducing two reciprocal transverse vector fields in free space without sources (his ${\bf K} = {\bf E}$, his ${\bf H} = -{\bf B}$). They have zero divergences, and satisfy coupled curl equations (Faraday and Amp\`{e}re-Maxwell) and separate wave equations. He then expresses the magnetic field in terms of the vector potential (his ${\bf \Gamma} = {\bf A}$):
\begin{quote}
``Wir wollen ${\bf H}$ ausw\"{a}hlen und das Potential mit ${\bf \Gamma}$ bezeichen, dann is zu setzen:
%\begin{equation}
\[ H_{x} \ =\ -\ \left(\frac{\partial \Gamma_{z}}{\partial y} \ -\ \frac{\partial \Gamma_{y}}{\partial z}\right);\ \  \ldots \ \]
%\end {equation}
\indent Damit wird  ${\bf \Gamma}$ noch nicht bestimmt; vor allen kommt in Betracht, dass der Werth von
%begin{equation}
\[ \frac{\partial \Gamma_{x}}{\partial x} + \frac{\partial \Gamma_{y}}{\partial y} +\frac{\partial \Gamma_{z}}{\partial z} \]
%end{equation}
willk\"{u}rlich bleibt; eine passende Verf\"{u}gung behalten wir uns vor.''
([div ${\bf \Gamma}$] remains arbitrary; we keep an appropriate choice in mind.)
\end{quote}
Wiechert then eliminates the magnetic field in favor of the vector potential in Faraday's law and finds the equivalent of our ${\bf E} = - {\bf \nabla}\Phi - \frac{\partial {\bf A}}{c \partial t}$, where $\Phi$ is the scalar potential (his $\phi = \Phi$). Eliminating the potentials in the source-free Coulomb's law and Amp\`{e}re-Maxwell equation leads him to:
%\begin{equation}
\[ \frac{\partial^{2} \phi}{\partial x^{2}}+ \frac{\partial^{2} \phi}{\partial y^{2}}+ \frac{\partial^{2} \phi}{\partial z^{2}}+ \frac{1}{V} \frac{\partial}{\partial t}\left(\frac{\partial \Gamma_{x}}{\partial x} + \frac{\partial \Gamma_{y}}{\partial y} +\frac{\partial \Gamma_{z}}{\partial z}\right) \ =\ 0 \ .\]
%end {equation}
and a corresponding equation for ${\bf \Gamma} $. Wiechert then states:
\begin{quote}
``Uber die Unbestimmtheit in ${\bf \Gamma}$ verf\"{u}gend stezen wir nun:
%\begin{equation}
\[\frac{\partial \phi}{\partial t}\ + \ V \left(\frac{\partial \Gamma_{x}}{\partial x} + \frac{\partial \Gamma_{y}}{\partial y} + 
\frac{\partial \Gamma_{z}}{\partial z} \right) \ = \ 0 \]
%\end{equation}
\end{quote}
This is Wiechert's version of the Loren(t)z condition, already quoted as Eq.(\ref{wiechert}). He then states that this relation simplifies the wave equations into the standard form and that the two wave equations, the Loren(t)z condition, and the definitions of the fields in terms of the potentials are equivalent to the Maxwell equations (for free fields).

Later in his paper, Wiechert adds charge and current sources to the equations and states the retarded solutions for his potentials. Other authors are cited, but not Lorenz.

Thirty-seven years after Lorenz and four years after Wiechert, Lorentz wrote two encyclopedia articles~\cite{lorentz3, lorentz1}, the second of which~\cite{lorentz1} contains on page 157 a discussion remarkably parallel (in reverse order) to that quoted earlier from Lorenz:
\begin{quote}
`` \ldots \ skalaren Potentials $\varphi$ und eines magnetischen Vektorpotentials ${\bf a}$ darstellen.  Es gen\"{u}gen diese Hilfsgr\"{o}ssen den Differentialgleichungen \\
\begin{eqnarray*}
(VII) \ \ \ \ \Delta \ \varphi - \frac{1}{c^{2}} \ddot{\varphi} & = & - \rho, \\
(VIII) \ \ \ \ \Delta \ {\bf a}- \frac{1}{c^{2}} \ddot{{\bf a}}  & = & - \frac{1}{c}\rho {\bf v},  \\
\end{eqnarray*}
und es ist
\begin{eqnarray*}
(IX) \ \ \ \ {\bf d} & = & -\frac{1}{c} \dot{{\bf a}} - grad \ \varphi, \\
(X) \ \ \ \ \ {\bf h} & = & rot \ {\bf a}.
\end{eqnarray*}
Zwischen den Potentialen besteht die Relation \\
\[ (2) \ \ \ div \ {\bf a} = - \frac{1}{c} \dot{\varphi}  \] \\
\end{quote}
Lorentz's (2) is the Loren(t)z condition displayed above as Eq.(\ref{lorentz}).  In an appendix in \emph{Theory of Electrons}(1909), he discusses gauge transformations and potentials that do not satisfy the Loren(t)z condition, but then states that he will always use potentials that satisfy Eq.(1). \\

While some may argue that Lorenz's 1967 paper was not a masterpiece of clarity, I would argue that his multifaceted approach makes clear that thirty-three years before Wiechert (a) he understood the arbitariness and equivalence of the different forms of potentials, and (b) he understood that Eq.(\ref{lorenz}) is a requirement for the retarded form of the Neumann (Loren(t)z gauge) potentials.\\

\section*{III.\ \ \  THE DIRAC DELTA FUNCTION}
\indent The second example is the Dirac delta function, popularized by Paul Adrien Maurice Dirac, British theoretical physicist, in his authoritative text \emph{The Principles of Quantum Mechanics}~\cite{dirac}, first published in 1930. There he introduces the (improper) impulse or delta function $\delta (x) $ in his discussion of the orthogonality and completeness of sets of basis functions in the continuum. His first definition is \\
\[ \ \delta (x) = 0 \ if \ x \neq 0 \ , \\ \int \delta (x) \ dx\ = \ 1. \]
Given the usefulness of the delta function in practical , if not rigorous, mathematics, it is not surprising that the delta function had ``discoverers'' before Dirac.  Oliver Heaviside, self-taught English electrical engineer, applied mathematician, and physicist, is arguably the person who should get credit for the introduction of the delta function. Thirty-five years before Dirac, in the March 15, 1895 issue of the British journal \emph{The Electrician}~\cite{heaviside}, he described his impulsive function in mathematical terms as \\
\[\ p{\bf 1} \ , where \ p \ =\ d/dt \ and \ {\bf 1} = \Theta (t) \]
Here $\Theta (t) $ is the Heaviside or step function ( $ \Theta (t) = 0$ for $t < 0 $, $\Theta (t) = 1 $ for $t > 0$, and $\Theta (0) = 1/2 $ ).

The origins of the delta function can be traced back to the early 19th century.~\cite{vanderpol}   Cauchy and Poisson, and later Hermite, used a function $D_{1}$:\\
\[ D_{1}(t) = \lim_{\lambda \rightarrow \infty}\ \frac{\lambda}{\pi(\lambda^{2} t^{2} +1)} \ , \] 
within double integrals in proof of the Fourier-integral theorem and took the limit $\lambda \rightarrow \infty $ at the end of the calculation. In the second half of the century Kirchhoff, Kelvin, and Helmholtz in other applications used similarly a function $ D_{2}(t) $: \\
\[ D_{2}(t) = \lim_{\lambda \rightarrow \infty} \ \frac{\lambda}{\surd \pi}exp(-\lambda^{2} t^{2} ) \]
While these sharply peaked functions presage the delta function, it was Heaviside and then Dirac who gave it explicit, independent status.

\subsection*{III.A  \ \ \  Oliver Heaviside (1850-1925)} 
Oliver Heaviside was born in London, England in 1850.  Illness in his youth left him partially deaf. Though an outstanding student, he left school at 16 to become a telegraph operator,with the help of his uncle Charles Wheatstone, wealthy inventor of the telegraph. Studying in his spare time, he began serious analysis of electromagnetism and publication in 1872 while working in Newcastle. Two years later, illness prompted him to resign his position to pursue research in isolation at his family home. There he conducted investigations of the skin effect, transmission line theory, and the beneficial influence of distributed inductance in preventing distortion and diminishing attenuation.  By 1885 Heaviside eliminated the potentials from Maxwell's theory and expressed it in terms of the four equations in four unknown fields, as we known them today. He, together with FitzGerald and Hertz, are credited with taking the mystery out of Maxwell's formulation. He is also responsible for introducing vector notation, independently of and contemporaneously with Gibbs; he discovered ``Poynting's theorem'' independently and found the ``Lorentz'' force of a magnetic field on a moving charged particle. In 1888-89 Heaviside evaluated the distorted patterns of the fields of a charge moving in vacuum and in a dielectric medium, the first influencing FitzGerald to think about a possible explanation of the Michelson-Morley experiment, and the second essentially a prediction of Cherenkov radiation. In the 1880s and 1890s he perfected and published his operational calculus for the benefit of engineers. In 1902, Kennelly and Heaviside independently proposed a conducting region in the upper atmosphere (Kennelly-Heaviside layer) as responsible for the long-distance propagation of telegraph signals around the earth.
	Self-educated and a loner, Heaviside jousted in print with ``the Cambridge mathematicians'' and was long ignored by the scientific establishment (with some notable exceptions).  He finally received recognition, becoming a Fellow of the Royal Society in 1891. He died in 1925.\\

\subsection*{III.B  \ \ \  Paul Adrien Maurice Dirac (1902-1984)}
Paul Dirac was born in Bristol, England in 1902 of a English mother and Swiss father. Educated in Bristol schools, including the technical college where his father taught French, Dirac studied electrical engineering at the University of Bristol, obtaining  his B. Eng. in 1921.  He decided on a more mathematical career and completed a degree in mathematics at Bristol in 1923.  He then went to St. John's College, Cambridge where he studied and published under the supervision of R. H. Fowler.  Fowler showed him the proofs of Heisenberg's first paper on matrix mechanics; Dirac noticed an analogy between the Poisson brackets of classical mechanics and the commutation relations of Heisenberg's theory. The development of this analogy led to his Ph.D. thesis,``Quantum mechanics,'' and publication in 1926 of his  mathematically consistent general theory of quantum mechanics in correspondence with Hamiltonian mechanics, an approach distinct from Heisenberg's and Schr\"{o}dinger's.  Dirac became a Fellow of St. John's College in 1927, the year he published his paper on the ``second'' quantization of the electromagnetic field. The relativistic equation for the electron followed in 1928. His treatise \emph{The Principles of Quantum Mechanics} (first edition, 1930) gave a masterful general formulation of the theory. Elected fellow of the Royal Society in 1930 and as Lucasian Professor of Mathematics at Cambridge in 1932, he shared the 1933 Nobel Prize in Physics with Schr\"{o}dinger ``for the discovery of new productive forms of atomic theory.'' Dirac made many other important contributions to physics - antiparticles, the quantization of charge through the existence of magnetic monopoles, the path integral approach ,... . He retired in 1969 and in 1972 accepted an appointment at Florida State University where he remained until his death in 1984.\\
\begin{figure} [h]
\centering
\includegraphics[width=3in]{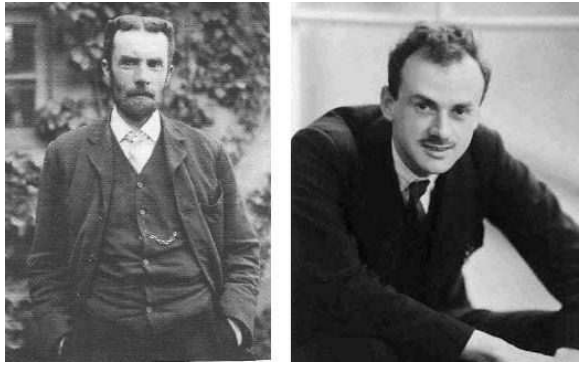}
\caption{(L-R) Oliver Heaviside, Paul A. M. Dirac}
\end{figure}

\subsection*{III.C  \ \ \  Textual Evidence}
From 1894 to 1898 Oliver Heaviside was publishing his operational calculus in \emph{The Electrician}.  In the March 15, 1895 issue he devoted a section to ``Theory of an Impulsive Current produced by a Continued Impressed Force.''~\cite{heaviside}  In it is the following partial paragraph:
\begin{quote}
	``We have to note that if $Q$ is any function of time, then $pQ$ is its rate of increase.   If, then, as in the present case, $Q$ is zero before and constant after $t = 0$, $pQ$ is then zero except when $t = 0$. It is then infinite.  But its total amount is $Q$. \emph{That is to say $p{\bf 1}$ means a function of $t$ which is wholly concentrated at the moment $t = 0$, of total amount $1$.  It is an impulsive function, so to speak.} The idea of an impulse is well known in mechanics,  and it is essentially the same here. Unlike the function $(p)^{1/2} {\bf 1}$, the function $p {\bf 1}$ does not involve appeal either to experiment or to generalised differentiation, but only involves the ordinary ideas of differentiation and integration pushed to their limit.''   [Emphasis added]
\end{quote}
In subsequent articles Heaviside  used his impulse function $p{\bf 1}$ extensively to treat various examples of the excitation of electrical circuits.

In Dirac's \emph{The Principles of Quantum Mechanics} (1930) he introduces the delta function and gives an extensive discussion of its properties and uses. In subsequent editions he alters the treatment somewhat; I quote from the third edition (1947)~\cite{dirac}: 
\begin{quote}
`` {\bfseries 15. The $\delta$ function} 
Our work in \S 10 led us to consider quantities involving a certain kind of infinity. To get a precise notation for dealing with these infinities, we introduce a quantity $\delta (x) $ depending on a parameter $x$ satisfying the conditions \\
\[  \int_{- \infty}^{\infty} \ \delta (x) \ dx \ = \ 1 \] \\
\[ \delta (x) \ = \ 0 \ \ for \ x \ \neq \ 0. \]  
To get a picture of $\delta (x) $, \emph{take a function of the real variable $x$ which vanishes everywhere except inside a small domain, of length $\epsilon$ say, surrounding the origin $x = 0$, and which is so large inside this domain that its integral over the domain is unity}. The exact shape of the function inside this domain does not matter, provided there are no unnecessarily wild variations (for example provided the function is always of order $\epsilon^{-1}$).  Then in the limit $\epsilon \rightarrow 0$ this function will go over into $\delta (x)$.''  [Emphasis added]
\end{quote}
A page later, Dirac gives an alternative definition: 
\begin{quote}
``An alternative way of defining the $\delta$ function is as the differential coefficient $\epsilon^{\prime} (x) $ of the function $\epsilon (x)$ given by
\begin{eqnarray*}
\epsilon (x) \ & = &  \ 0 \ \ \ (x < 0)  \\
	          & = &  \ 1 \ \ \ (x > 1) 
\end{eqnarray*}
We may verify that this is equivalent to the previous definition . . .''
\end{quote}
This definition is explicitly Heaviside's definition ($\epsilon (x) = {\bf 1},   \epsilon^{\prime} = p {\bf 1}$).  And the descriptions in words are strikingly similar, but what else could they be?\\

\section*{IV. \ \ \ SCHUMANN RESONANCES OF THE EARTH-IONOSPHERE CAVITY}
The third example concerns ``Schumann resonances,'' the extremely low frequency (ELF) modes of electromagnetic waves in the resonant cavity formed between the conducting earth and ionosphere.  Dimensional analysis with the speed of light $c$ and the circumference of the earth $2 \pi R$ gives an order of magnitude for the frequency of the lowest possible mode, $\nu_{0} = O(c/2 \pi R) = 7.45 \ Hz$.~\cite{height}. The spherical geometry, with Legendre functions at play, leads to a series of ELF modes with frequencies, $ \nu (n) = \surd n(n+1) \ \nu_{0}$ for the cavity between two perfectly conducting spherical surfaces. It was perhaps J. J. Thomson who first solved for these modes (1893), with many others subsequently. It was Winfried Otto Schumann, a German electrical engineer, who in 1952 applied the resonant cavity model to the earth-ionosphere cavity,~\cite{schumann} although others before him had used wave-guide concepts.~\cite{besser}  Because the earth and especially the ionosphere are not very good conductors, the resonant lines are broadened and lowered in frequency, but still closely following the Legendre function rule, with $\nu (n) \approx 5.8 \surd n(n+1) \ Hz \ =\ 8, 14, 20, 26, ... \ Hz$. In a series of papers from 1952 to 1957, Schumann discussed damping, the power spectrum from excitation by lightning, and other aspects. Since their first clear observation in 1960, the striking resonances have been studied extensively.~\cite{besser, jackson}  \\

Although Schumann can be said to have initiated the modern study of extreme ELF propagation and many have been occupied with the peculiarities of long-distance radio transmission since Kennelly and Heaviside, two names emerge as earlier students of at least the lowest ELF mode around the earth. Those names and dates are Nicola Tesla, Serbian-American inventor, physicist, and engineer, (1905) and George Francis FitzGerald, Irish theoretical physicist, (1893). Indeed, there are those that claim that Tesla actually observed the resonance. 

\subsection*{IV.A \ \ \ George Francis FitzGerald (1851-1901)}
George Francis Fitzgerald was born in 1851 near Dublin and home-schooled; his father was a minister and later a bishop in the Irish Protestant Church.  He studied mathematics and science at the University of Dublin, receiving his B.A. in 1871. For the next six years he pursued graduate studies, becoming a fellow of Trinity College, Dublin in 1877. He served as college tutor and as a member of the Department of Experimental Physics until 1881 when he was appointed Professor of Natural and Experimental Philosophy, University of Dublin. FitzGerald's researches were largely but not exclusively in optics and electromagnetism.  Working out the amount of radiation emitted by discharging circuits in 1883, he foresaw the possibility  of Hertz's experiments; in 1889 he had the intuition that a length contraction proportional to $v^{2}/c^{2}$ in the direction of motion could explain the null effect of the Michelson-Morley experiment (FitzGerald-Lorentz contraction). He was elected Fellow of the Royal Society in 1883. A model professional citizen, FitzGerald  served as officer in scientific societies, as external examiner in Britain, and on Irish committees concerned with national education. He died in 1901 at the early age of 49. \\
\begin{figure} [htp]
\centering
\includegraphics[width=1.5in]{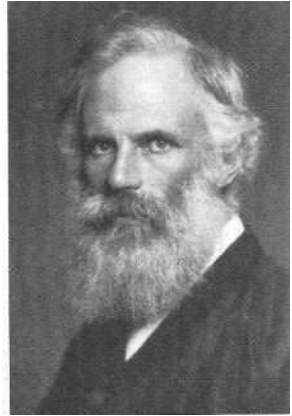}
\caption{George F. FitzGerald }
\end{figure}

\subsection*{IV.B \ \ \ Nikola Tesla (1856-1943)}
Nikola Tesla was born in Smiljan, Croatia in 1856 of Serbian parents. He studied electrical engineering at the Technical University in Graz, Austria and at Prague University. He worked in Paris as an engineer, 1882-83, and then in 1884 emigrated to the US where he worked for a short time for Thomas Edison. But in May 1885 Tesla switched to work for Edison's competitor, George Westinghouse, to whom Tesla sold his patent rights for a-c dynamos, polyphase transformers, and a-c motors. Later he set up an independent laboratory to pursue his inventions. He became a US citizen in 1891, the year he invented the tesla coil. For six or seven months in 1899-1900 Tesla was based in Colorado Springs where he speculated about terrestrial standing waves and conducted various startling experiments such as man-made lightning bolts up to 40 meters in length. In 1900 he moved to Long Island where he began to built a large tower for long-distant transmission of electromagnetic energy. In his lifetime he had hundreds of patents. Although in later life he was discredited for his wild claims and died impoverished in 1943, he is recognized as the father of the modern a-c high-tension power distribution system used worldwide. 

\subsection*{IV.C \ \ \ Winfried Otto Schumann (1888-1974)}
Winfried Otto Schumann was born in T\"{u}bingen, Germany in 1888, the son of a physical chemist. He studied electrical engineering at the Technische Hochschule in Karlsruhe, earning his first degree in 1909 and his Dr.-Ing. in 1912. He worked in electrical manufacturing until 1914; during World War I he served as a radio operator. In 1920 Schumann was appointed as Associate Professor of Technical Physics at the University of Jena. In 1924 he became Professor for Theoretical Electrical Engineering, Technische Hochschule, Munich (now the Technical University) where he remained until retirement, apart from a year (1947-48) at the Wright-Patterson Air Force Base in Ohio. Schumann's early research was in high-voltage engineering.  In Munich, for 25 years his interests were in plasmas and wave propagation in them. Then from 1952 to 1957, as already noted, he worked on ELF propagation in the earth-ionosphere cavity. Later, into retirement after 1961, his research was in the motion of charges in low-frequency electromagnetic fields.~\cite{besser} Schumann died in 1974 at the age of 86.\\
\begin{figure} [htp]
\centering
\includegraphics[width=3in]{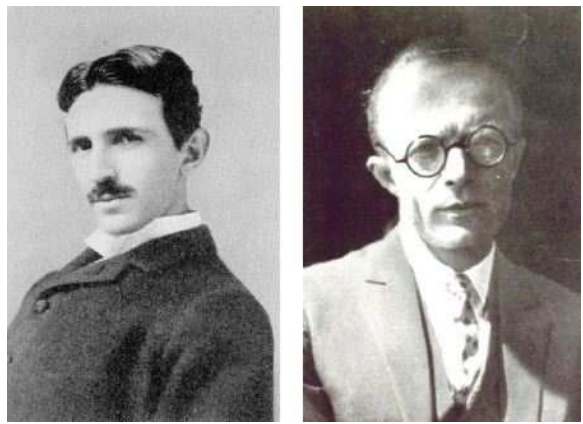}
\caption{(L-R) Nikola Tesla, Winfried O. Schumann}
\end{figure} 
\subsection*{IV.D \ \ \ Textual Evidence }
In 1900 Tesla filed a patent application entitled, ``Art of transmitting electrical energy through the natural mediums.'' The United States Patent Offioce granted him the Patent No. 787,412 on April 18, 1905.~\cite{tesla}  To convey the thrust of Tesla's reasoning regarding the transmission of very low frequency electromagnetic energy over the surface of the earth, I quote important excerpts. \\
\begin{quote}
``... For the present it will be sufficient to state that the planet behaves like a perfectly smooth or polished conductor of inappreciable resistance with capacity and self induction uniformly distributed along the axis of symmetry of wave propagation and transmitting slow electrical oscillations without sensible distortion or attenuation.  ....'' 
\end{quote}
Tesla treats the earth as a perfectly conducting sphere in infinite space. He does not know of the ionosphere or conduction in the atmosphere. \\
\begin{quote}
``First. The earth's diameter passing through the pole should be an odd multiple of the quarter wave length - that is, of the ratio of the velocity of light - and four times the frequency of the currents.''
\end{quote} 
Here Tesla seems to be thinking of propagation through the earth. His description translates into a frequency of oscillation $\nu (n) = (2n+1)c/8R \approx 5.9 (2n+1) \ Hz $.
\begin{quote}
``.... To give an idea, I would say that the frequency should be smaller than twenty thousand per second, though shorter waves might be practicable.  The lowest frequency would appear to be six per second, ....''
\end{quote}
Tesla is thinking of power transmission, not radiation into space, and so is keeping the frequency down, $6 \ Hz $ being his minimum. \\
\begin{quote}
``Third. The most essential requirement is, however, that irrespective of frequency the wave or wave-train should continue for a certain interval of time, which I estimated to be not less than one twelfth or probably 0.08484 of a second and which is taken passing to and returning from the region diametrically opposite the pole over the earth's surface with a mean velocity of about four hundred and seventy-one thousand two hundred and forty kilometers per second.''
\end{quote}
The stated speed, given with such accuracy, is $\pi/2 $ times the speed of light \emph{c}. It makes the time for a pulse to travel over the surface from pole to pole equal to the time taken at speed \emph{c} along a diameter. It would be  natural to wish a pulse to have a certain duration if resonant propagation were envisioned, but the special significance of 0.08484 seconds is puzzling.  Equating the surface time to the diameter time seems to tie back to his use of the diameter to find the frequencies. \\

That Tesla had ideas about low frequency electromagnetic modes encompassing the whole earth is clear. But he did not envision the conducting layer outside the earth's surface that creates a resonant cavity. There is no evidence that he ever observed propagation around the earth. And a decade earlier, FitzGerald discussed the phenomenon realistically. \\ 

In September 1893 FitzGerald presented a paper at the annual meeting of the British Association for the Advancement of Science.~\cite{fitzgerald1}  An anonymous correspondent gave a summary of FitzGerald's talk in \emph{Nature}.~\cite{nature}  I quote first from the Report of the British Association, which seems to be an abstract, submitted in advance of the meeting:
\begin{quote}
``Professor J. J. Thomson and Mr. O. Heaviside have calculated the period of vibrations of a sphere alone in space and found it about 0.59 second. The fact that the upper regions of the atmosphere conduct makes it possible that there is a period of vibration due to the vibrations similar to those on a sphere surrounded by a concentric spherical shell.  In calculating this case it is not necessary to consider propagation in time for an approximate result, . . . . The value of the time of vibration obtained by this very simple approximation is
\[ T \ = \ \pi \sqrt{\frac{2K\mu a^2 b^2 log(a/b)}{a^2 - b^2}} \]
Applying this to the case of the earth with a conducting layer at a height of 100 kilometres (much higher than is probable) it appears that a period of vibration of about one second is possible. A variation in the height of the conducting layer produces only a small effect upon this if the height be small compared to the diameter of the earth. . . . .''
\end{quote}
\noindent FitzGerald's mention of one second is a bit curious, but may be a typographical error. In the limit of $ b\rightarrow a $, his formula yields $ T = \pi a/c \ \approx \ 1/15 \  Hz^{-1} $, a value that is off by just $\sqrt{2} $ from the correct $ T = 1/10.6 \  Hz^{-1} $ for perfect conductivity. \\

In the account of the BA meeting in the September 28, 1893 issue of \emph{Nature}, the reporter notes that `` Professor G. F. FitzGerald gave an interesting communication on `The period of vibration of disturbances of electrification of the earth.' ''  He notes the following points made by FitzGerald:
\begin{quote}
1. `` . . . . the hypothesis that the Earth is a conducting body surrounded by a non-conductor is not in accordance with the fact. Probably the upper regions of our atmosphere are fairly good conductors.''
2. `` . . . . we may assume that during a thunderstorm the air becomes capable of transmitting small disturbances.''
3. ``If we assume the height of the region of the aurora to be 60 miles or 100 kilometres, we get a period of oscillation of 0.1 second.''
\end{quote}
\noindent Now the period of vibration is correct. \\

It is clear that in 1893 FitzGerald had the right model, got roughly the right answer for the lowest mode, and had the prescience to draw attention to thunderstorms, the dominant method of excitation of Schumann resonances.\\

\section*{V. \ \ \ WEIZS\"{A}CKER-WILLIAMS  METHOD OF VIRTUAL QUANTA}
My fourth example is the Weizs\"{a}cker-Williams  method of virtual quanta, a theoretical approach to inelastic collisions of charged particles at high energies in which the electromagnetic fields of one of the particles in the collision are replaced by an equivalent spectrum of virtual photons. The process is then described in terms of the inelastic collisions of photons with the ``target.''  C. F. von Weizs\"{a}cker and E. J. Williams were both at Niels Bohr's Institute in Copenhagen in the early 1930s when the validity of quantum electrodynamics at high energies was in question. Their work~\cite{weiz, williams1, williams2} in 1934-35  played an important role in assuaging those fears. The concept has found wide and continuing applicability in particle physics, beyond purely electrodynamic processes.\\

But they were not the first to use the method. Ten years earlier, in 1924, even before the development of quantum mechanics, Enrico Fermi discussed the excitation and ionization of atoms in collisions with electrons and energy loss using what amounts to the method of virtual quanta.~\cite{fermi}  Fermi was focused mainly on nonrelativistic collisions; a key aspect of the work of  Weizs\"{a}cker and Williams, the appropriate choice of inertial frame in which to view the process, was missing. Nevertheless, the main ingredient, the equivalent spectrum of virtual photons to replace the fields of a charged particle, is Fermi's invention.

\subsection*{V.A \ \ \ Enrico Fermi(1901-1954)}
One of the last ``complete'' physicists, Enrico Fermi was born in Rome. His father was a civil servant. At an early age he took an interest in science, especially mathematics and physics. He received his undergraduate and doctoral degrees from the Scuola Normale Superiore in Pisa. After visiting G\"{o}ttingen and Leiden in 1924, he spent 1925-26 at the University of Florence where he did his work on what we call the Fermi-Dirac statistics of identical spin 1/2 particles. He then took up a professorship in Rome where he remained until 1938. He soon was leading a powerful experimental group that included Edoardo Amaldi, Bruno Pontecorvo, Franco  Rasetti, and Emilio Segr\`{e}. Initially, their work was in atomic and molecular spectroscopy, but with the discovery of the neutron in 1932 the group soon switched to nuclear transmutations induced by slow neutrons and became preeminent in the field. Stimulated by the Solvay Conference in Fall 1933, where the neutrino hypothesis was sharpened, Fermi quickly created his theory of beta decay in late 1933/early 1934. He was awarded the Nobel Prize in Physics for the nuclear transmutation work in 1938. He took the opportunity to emigrate from Sweden to the U.S. in December that year, just as the news of the discovery of neutron-induced nuclear fission became public. Initially at Columbia, Fermi moved to the University of Chicago where, once the Manhattan District was created, he was in charge of construction of the first successful nuclear reactor (1942). Later he was at Los Alamos. After the war he returned to Chicago to build a synchrocyclotron powerful enough to create pions and permit study of their interactions. In his nine years at Chicago he mentored a very distinguished group of Ph.D. students, five of whom later became Nobel Laureates. \\ 
\begin{figure} [htp]
\centering
\includegraphics[width=1.5in]{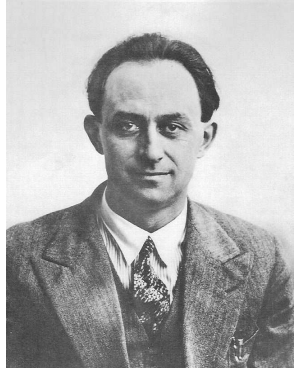}
\caption{Enrico Fermi (Rome years)}
\end{figure} 
\subsection*{V.B \ \ \ Carl Friedrich von Weizs\"{a}cker (1912-2007)}
Carl Friedrich von Weizs\"{a}cker, son of a German diplomat, was born in Kiel, Germany. From 1929 to 1933 he studied physics, mathematics, and astronomy in Berlin (with Schr\"{o}dinger), G\"{o}ttingen (briefly, with Born), and Leipzig, where he was Heisenberg's student, obtaining his Ph.D. in 1933. He was at Bohr's Institute, 1933-34, where he did the work we discuss here. In the 1930s his research was in nuclear physics and astrophysics. Notable was his work on energy production in stars, done contemporaneously with Hans Bethe. During World War II he joined Heisenberg in the German atomic bomb effort. He is credited with realizing that plutonium would be an alternative to uranium as a fuel for civilian energy production. After the war, he was the spokesman for the view that the German project was aimed solely at building a nuclear reactor, not a bomb. 
In 1946 he went to the Max Planck Institute in G\"{o}ttingen. His interests broadened to the philosophy of science and technology and their interactions with society. He became Professor of Philosophy at the University of Hamburg in 1957.
Then in 1970 until his retirement in 1980, he was the director of a Max Planck Institute for the study of the living conditions of the scientific-technological world. The later Weizs\"{a}cker was a prolific author on the philosophy of science and society, and an activist on issues of nuclear weapons and public policy.\\
\begin{figure} [h]
\centering
\includegraphics[width=3in]{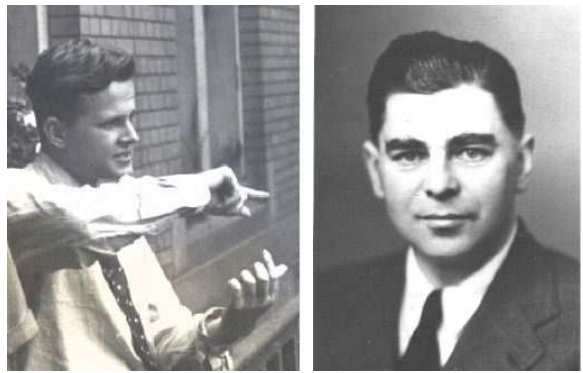}
\caption{(L-R) Carl .F. von Weizs\"{a}cker, Evan J. Williams}
\end{figure} 
\subsection*{V.C \ \ \ Evan James Williams (1903-1945)}
Evan James Williams was born in Cwmsychpant, Wales, and received his early education at Llandysul County School where he excelled in literary and scientific pursuits. A scholarship student at the University of Wales, Swansea, he graduated with a M.Sc. in 1924.  Williams then studied for his Ph.D. at the University of Manchester under W. L. Bragg; a further Ph.D. was earned at Cambridge in 1929 and a Welsh D.Sc. a year later. His research was in both experiment and theory. Nuclear and cosmic ray studies led to theoretical work on quantum mechanical calculations of atomic collisions and energy loss. He spent 1933 at Bohr's Institute in Copenhagen where he worked in a loose collaboration with Bohr and Weizs\"{a}cker. He then held positions at Manchester and Liverpool before accepting the Chair of Physics at University of Wales, Aberystwyth in 1938. Elected fellow of the Royal Society in 1939, a year later Williams and G. E. Roberts used a cloud chamber to make the first observation of muon decay. He served in the Air Ministry and Admiralty during World War II. His career was cut short in 1945 at age 42. In his review of the penetration of charged particles in matter published in 1948, Niels Bohr laments that the review had originally been intended to be a collaboration with Williams. \\

\subsection*{V.D \ \ \ Textual Evidence }
A newly minted Ph.D. in 1924, Enrico Fermi addressed the excitation of atoms in collisions with electrons and the energy loss of charged particles in a novel way.~\cite{fermi}  The abstract of his paper is
\begin{quote}
``Das elektrische Feld eines geladenen Teilchens, welches an einem Atom vorbeifliegt, wird harmonisch zerlegt, und mit dem elektrischen Feld von Licht mit einer passenden Frequenzverteilung verglichen.  Es wird angenommen, dass die Wahrscheinlichkeit, dass das Atom vom vorbeifliegenden Teilchen angeregt oder ionisiert wird, gleich ist der Wahrscheinlichkeit f\"{u}r die Anregung oder Ionisation durch die \"{a}quivalente Strahlung.  Diese Annahme wird angewendet auf die Anregung durch Elektronenstoss und auf die Ionisierung und Reichweite der $\alpha$-Strahlen.
\end{quote}
A rough literal translation is
\begin{quote}
``The electric field of a charged particle that passes by an atom, when decomposed into harmonics, is equivalent to the electric field of light with an appropriate frequency distribution. It will be assumed that the probability that an atom will be excited or ionized by the passing particle is equal to the probability for excitation or ionization through the equivalent radiation.  This hypothesis will be applied to the excitation through electron collisions and to the ionizing power and range of $\alpha$-particles.''
\end{quote}
That first sentence describes a key ingredient of the Weizs\"{a}cker-Williams method of virtual quanta. Because he was working before quantum mechanics had emerged, Fermi had to use empirical data for the photon-induced ionization and excitation of atoms to fold with the equivalent photon distribution.  Explicitly, Fermi's expression for the probability of inelastic collision of a charged particle and an atom, to be integrated over equivalent photon frequencies and impact parameters of the collision, is \\
\[ dP \ = \ \frac{J(\nu) \ \alpha(\nu)}{h\nu} \ d\nu \ 2\pi b db \]
where $J(\nu)$ is the nonrelativistic limit of the equivalent photon flux density and $\alpha(\nu)$ is the photon absorption coefficient.  For K-shell ionization, for example, an approximate form is \\
\[ \alpha(\nu) \ = \frac{\ H \ Z^{4}}{\nu^{3}} \ \Theta(\nu - \nu_{0})  \]
where $\nu_{0}$ is the K-shell threshold and \emph{H} is an empirical constant.\\

Nine years later, E. J. Williams, in his own work on energy loss,~\cite{williams3}  discussed the limitations of Fermi's work in the light of proper quantum mechanical calculations of the absorption of a photon by an atom. \emph{A year later}, in Part III of his letter to the \emph{Physical Review},~\cite{williams1} after citing his 1933 approach to collisional energy loss using semi-classical methods, Williams said,
\begin{quote}
`` . . . . Practically the same considerations apply to the formula of Heitler and Sauter for the energy lost by an electron in radiative collisions with an atomic nucleus. C. F. v. Weizs\"{a}cker and the writer, in calculations shortly to appear elsewhere, show that this formula may readily be derived by considering, in a system $S'$ where the electron is initially at rest, the scattering by the electron of the harmonic components in the Fourier spectrum of the perturbing force due to the nucleus (which, in $S'$, is the moving particle). The calculations show that practically all the radiative energy loss comes from the scattering of those components with frequencies $\sim mc^{2}/h$, and also that Heitler and Sauter's formula is largely free from the condition $Ze^{2}/\hbar c << 1$, which generally has to be satisfied in order that Born's approximation (used by H and S) may be valid . . . ''
\end{quote}
The virtual quanta of the fields of the nucleus passing an electron in its rest frame $S'$ are Compton-scattered to give bremsstrahlung.  Hard photons in the lab come from $\hbar \omega \sim mc^{2}$ in the rest frame $S'$. \\ 

Weizs\"{a}cker~\cite{weiz} used the equivalent photon spectrum together with the Klein-Nishina formula for Compton scattering to show that the result was identical to the familiar Bethe-Heitler formula for bremsstrahlung. In a long paper published in 1935 in the Proceedings of the Danish Academy,~\cite{williams2} Williams presented a more general discussion,``Correlation of certain collision problems with radiation theory,'' with the first reference being to Fermi.   Weizs\"{a}cker and Williams exploited special relativity to show that in very high energy radiative processes the dominant energies are always of order of the light particle's rest energy when seen in the appropriate reference frame. The possible failure of quantum electrodynamics at extreme energies, posited by Oppenheimer and others, does not occur. The apparent anomalies in the cosmic rays were in fact evidence of then unknown particles (muons).\\

Fermi started it; Williams obviously knew of Fermi's virtual photons; he and Weizsa\"{a}cker chose the right rest frames for relativistic processes. The ``Weizs\"{a}cker-Williams  method of virtual quanta'' continues to have wide and frequent applicability.\\

\section*{VI. \ \ \ BMT EQUATION FOR SPIN MOTION IN ELECTROMAGNETIC FIELDS}
In 1959 Valentine Bargmann, Louis Michel, and Valentine Telegdi published a short paper~\cite{bmt} on the behavior of the spin polarization of a charged particle with a magnetic moment moving relativistically in fixed, slowly varying (in space) electric and magnetic fields. The equation, known colloquially as the BMT equation in a pun on a New York City subway line, finds widespread use in the accelerator physics of high energy electron-positron storage rings. While the authors cite some earlier specialized work, notably by J. Frenkel and H. A. Kramers, they do not cite the true discoverer and expositor of the general equation.
\indent In the April 10, 1926 issue of \emph{Nature} Llewellyn H. Thomas published a short letter~\cite{thomas1}  explaining and eliminating the puzzling factor of two discrepancy between the atomic fine structure and the anomalous Zeeman effect, a paper that is cited for what we know as the ``Thomas factor'' (of 1/2). Thomas, then at Bohr's Institute, had listened before Christmas 1925 to Bohr and Kramers arguing over Goudsmit and Uhlenbeck's proposal that the electron had an intrinsic spin. They concluded that the factor of two discrepancy was the idea's death knell. Thomas suggested that a relativistic calculation should be done and did the basic calculation over one Christmas weekend in 1925.~\cite{thomas2} He impressed Bohr and Kramers enough that they urged the letter to \emph{Nature}. Then Thomas elaborated in a detailed 22-page paper~\cite{thomas3} that appeared in   
a January 1927 issue of Philosophical Magazine and is not widely known. It is this paper that contains all of BMT and more.\\

\subsection*{VI.A \ \ \ Llewellyn Hilleth Thomas(1903 - 1992)}
Born in London, England, Llewellyn Hilleth Thomas was educated at the Merchant Taylor School and Cambridge University, where he received his B.A. in 1924. He began research under the direction of R. H. Fowler who promptly went to Copenhagen, leaving Thomas to his own devices. In recompense Fowler arranged for Thomas to spend the year 1925-26 at Bohr's Institute where, among other things, he did the celebrated (and neglected) work described here. On his return to Cambridge 
he was elected a Fellow of Trinity College while still a graduate student. He received his Ph.D. in 1927.\\
\indent In 1929 Thomas emigrated to the US, to Ohio State University, where he served for 17 years. Notable while at Ohio State was his invention in 1938 of the sector-focusing cyclotron, designed to overcome the effects of the relativistic change in the cyclotron frequency. During World War II he worked at the Aberdeen Proving Ground. From 1946 to 1968 he was at Columbia University and the IBM Watson Laboratory. There he did research on computing and computers, including invention of a version of the magnetic core memory. He retired from Columbia and IBM in 1968 to become University Professor at North Carolina State University until a second retirement in 1976. He was a member of the National Academy of Sciences.\\
\begin{figure}[h]
\centering
\includegraphics[width=1.5in]{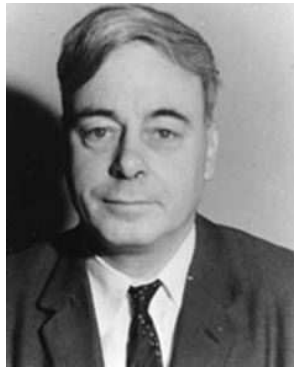}
\caption{Llewellyn H. Thomas}
\end{figure} 
\subsection*{VI.B \ \ \ Valentine Bargmann (1908 - 1989)}
Valentine Bargmann was born and educated in Berlin. He attended the University of Berlin from 1925 to 1933. Then with the rise of Hitler, he moved to the University of Zurich under Gregor Wentzel for his Ph.D. Soon after, he emigrated to the US. He served as an assistant to Albert Einstein at the Institute for Advanced Study where he collaborated with Peter Bergmann.  During World War II Bargmann worked with John von Neumann on shock wave propagation and numerical methods. In 1946 he joined the Mathematics Department at Princeton University. There he did research on mathematical physics topics, including the Lorentz group, Lie groups, scattering theory, and Bargmann spaces, collaborating famously with Eugene Wigner in 1948 on relativistic equations for particles of arbitrary spin, and of course with M and T of BMT. He was awarded several prizes and elected a member of the National Academy of Sciences.\\

\subsection*{VI.C \ \ \ Louis Gabriel Michel (1923 - 1999)}
Louis Michel was born and grew up in Roanne, Loire, France. He entered Ecole Polytechnique in 1943 and, after military service, joined the ``Corps des Poudres,'' a governmental basic and applied research institution, and was assigned back to Ecole Polytechnique to do cosmic ray research. He was sent to work with Blackett in Manchester, but in 1948 began theoretical work with Leon Rosenfeld. He completed his Paris Ph.D. in 1950 on weak interactions, especially the decay spectrum of electrons from muon decay and showed its dependence (ignoring the electron's mass) on only one parameter, known as the ``Michel parameter.''  Michel spent time in Copenhagen in the fledgling CERN theory group and at the Institute for Advanced Study in Princeton before returning to France. He held positions in Lille, Orsay, and Ecole Polytechnique, and finally from 1962 at the Institut des Hautes Etudes Scientifiques at Bures-sur-Yvette. A major part of Michel's research concerned spin polarization in fundamental processes, of great importance after the discovery of parity non-conservation in 1957, with the BMT paper somewhat related. Later research spanned strong interactions and G-parity, symmetries and broken symmetries in particle and condensed matter physics, and mathematical tools for crystals and quasi-crystals. Michel was a leader of French science, President of the French Physical Society, member of the French Academy of Sciences, and recipient of many other honors. \\

\subsection*{VI.D \ \ \ Valentine Louis Telegdi (1922 - 2006)}
Although born in Budapest, Valentine Louis Telegdi spent his minority moving all around Europe with his family, a likely explanation for his fluency in numerous languages. The family was in Italy in the 1940s. In 1943 they finally found refuge from the war in Lausanne, Switzerland where Telegdi studied at the University.  In 1946 Telegdi began graduate studies in nuclear physics at ETH Z\"{u}rich in Paul Scherrer's group. He came to the University of Chicago in the early 1950s. There he exhibited his catholic interests in research. Noteworthy was a paper in 1953 with Murray Gell-Mann on charge independence in nuclear photo-processes.  His name is associated with a wide variety of important measurements or discoveries: $G_{A}/G_{V}$, the ratio of axial-vector to vector coupling in nuclear beta decay; $(g-2)_{\mu}$, measurement of the anomalous magnetic moment of the muon; $K_{s}$ regeneration; muonium, an atomic-like bound state of a positive muon and an electron; and numerous others. Perhaps the best known work is the independent discovery of parity violation in the pion-muon decay chain, published in early 1957, with Jerome Friedman.\\ 
\indent  In 1976 Telegdi moved back to Switzerland to take up a professorship at ETH Z\"{u}rich, with research and advisory roles at CERN. In retirement he spent time each year at CalTech and UCSD. Among his many honors were memberships in the US National Academy of Sciences and the Royal Society, and, in 1991, co-winner of the Wolf Prize.\\
\begin{figure} [h]
\centering
\includegraphics[width=4.5in]{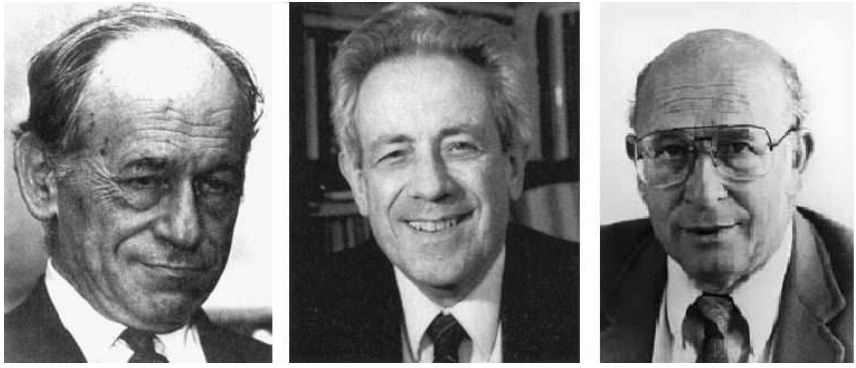}
\caption{(L-R) Valentine Bargmann, Louis Michel, Valentine Telegdi}
\end{figure} 
\subsection*{VI.E \ \ \ Textual Evidence }
To show the close parallel between Thomas's work and the BMT paper 32 years later, we quote significant equations from both in facsimiles of the original notation.  In both Thomas's 1927 paper and the BMT paper of 1959 the motion of the charged particle is described by the Lorentz force equation, with no contribution from the action of the fields $F^{\mu \nu}$ on the magnetic moment. In Thomas's text the Lorentz force equation reads \\
\[ \frac{d}{ds}(m\frac{dx^{\mu}}{ds}) \ = \ \frac{\epsilon}{c} F^{\mu}\ _{\nu} \frac{dx^{\nu}}{ds}  \ \ \ \ \ \ \ \ (1.73) \]\\
Here $ x^{\mu} $ is the particle's space-time coordinate and $s$ is its proper time.  In BMT's notation the equation reads more compactly as \\
\[du/d\tau \ = \ (e/m)F \cdot u \ \ \ \ \ \ \ \ \ \ \ \ \ (5)  \]
where $u$ is the particle's 4-velocity, and now $\tau$ is the proper time.
\indent For the spin polarization motion, we quote first  the BMT equation:\\
\[ ds/d\tau \ = \ (e/m)[(g/2)F \cdot s \ + \ (g/2 - 1)(s \cdot F \cdot u)u \ ] \ \ \ \ \ (7) \]\\
Here $ s (= s^{\mu}) $ is the particle's 4-vector of spin angular momentum and $ g $ is the g-factor of the particle's magnetic moment, $ \vec{\mu} = g e \hbar \vec{s}/2mc$.
For spin motion Thomas used both a spin 4-vector $w^{\mu}$ and an antisymmetric second-rank tensor $ w^{\mu \nu} $. Here is Thomas's equivalent to the spatial part of BMT's spin equation, as he wrote it out explicitly.  His $\beta $ is what is usually called the relativistic factor $\gamma$; his $\lambda = (g/2)e/mc$:\\
\begin{eqnarray*}
\frac{d{\bf w}}{ds} \ &=& \ \left[\{\frac{e}{mc}+ \beta (\lambda - \frac{e}{mc}) \}{\bf H}+\frac{(1-\beta)}{{\bf  v}^{2}}(\lambda - \frac{e}{mc})({\bf H \cdot v}){\bf v} \right. \\
				 & & \: + \left.(\frac{e}{mc^{2}}\frac{\beta^{2}}{1+\beta} - \frac{\lambda \beta}{c})[{\bf v \times E}]\:\right]\ {\bf \times \ w} \ \ \ \ \ \ \ \ \ \  (4.121)
\end{eqnarray*}
Thomas then says, 
\begin{quote}
``This last is considerably simpler when $\lambda = e/mc $, when it takes the form \\
\[\frac{d{\bf w}}{ds} = \left\{\frac{e}{mc}{\bf H} \ - \ \frac{e}{mc^{2}}\frac{\beta}{1+\beta}[{\bf  v \times E}]\:\right\}{\bf \times \ w} \ \ \ \ \ \ \ \ \ (4.122) \]
``In this case in the same approximation,\\
\[ \frac{dw^{\mu}}{ds} = \frac{e}{mc}\ F^{\mu}_{\nu}w^{\nu} \ \ \ \ \ \ \ \ \ \ \ \ \ \ \ \ \ \ \ \ \ \ \ \ \ \ \  (4.123) \]
\indent and \\
\[ \frac{dw_{\mu \nu}}{ds} = \frac{e}{mc}\{F^{\sigma}_{\nu}w_{\sigma \mu}\ -\ F^{\sigma}_{\mu} w_{\sigma \\nu} \ \} \ \ \ \ \ \ \ \ \ \ \ \ (4.124) \]
``The more complicated forms when $\lambda \neq e/mc$ involving ${\bf v}$ explicitly on the right-hand side can be found easily if required.''
\end{quote}
\noindent Compare Thomas's (4.123) for $g = 2$ with the corresponding BMT eqnation, $ds/d\tau = (e/m) \ F\cdot s$. Clearly Thomas felt that his explicit general form (4.121) for arbitary g-factor (arbitrary $\lambda$) was more use than a compact 4-vector form such as the BMT equation (7).\\

In 1926-27 Thomas was concerned about atomic physics; his focus was on the ``Thomas factor'' in the comparison of the fine structure and the anomalous Zeeman effect in hydrogen. Bargmann, Michel, and Telegdi focused on relativistic spin motion and how electromagnetic fields changed transverse polarization into longitudinal polarization and vice versa, with application to the measurement of the muon's g-factor in a storage ring. But it was all in Thomas's 1927 paper, 32 years earlier. \\

\section*{VII. \ \ \ CONCLUSIONS}
These five examples from physics illustrate the different ways that inappropriate attributions are given for significant contributions to science. 
Ludvig Vladimir Lorenz lost out to the homophonous Dutchman (and Emil Wiechert) largely because he was in some sense before his time and was a Dane who published an appreciable part of his work in Danish only. He died in 1891, just as Lorentz was most productive in electromagnetic theory and applications. By 1900 Lorenz's name had virtually vanished from the literature. Recently a move is underway (in which the author is a participant) to recognize Lorenz for the ``Lorentz'' condition and gauge; H. A. Lorentz has many other achievements attributed justly to him.\\

In the physics literature the Dirac delta function often goes without attribution, so common is its usage. But when a name is attached, it is Dirac's, not Heaviside's.  The reason, I think, is that in the 1930's and 1940's Dirac's book on quantum mechanics  was a standard and extremely influential.  With almost no references in the book, it is not surprising the Dirac did not cite Heaviside, even though, as an electrical engineer, he surely was aware of Heaviside's operational calculus and his impulse function.  Those learning quantum mechanics then (and now) would be only dimly or not at all aware of who was Oliver Heaviside. Scientists look forward, not back; 35 or 40 years is a lifetime or two. I can hear the voices saying: If some electrical engineer used the same concept in the 19th century, so be it. But our delta function is Dirac's. \\

Schumann resonances are a different case. For some years, now enhanced by the Internet, a vocal minority have trumpeted Tesla's discovery of the low-frequency electromagnetic resonances in the Earth-ionosphere cavity. I believe that claim is incorrect, but Tesla was a genius in many ways. It is not surprising that his discussion of resonances around the Earth in his 1905 patent might be interpreted by some as a prediction or even discovery of Schumann resonances. The interesting aspect is that it was a theoretical physicist, not an electrical engineer, who first discussed the Earth-ionosphere cavity, and in an insightful way.  FitzGerald, in 1893, was indeed well before the appropriate time.  And a talk to the British Association, followed by brief mention in a column in \emph{Nature}, is not a prominent literature trail for later scientists. Schumann may be forgiven for not citing FitzGerald, even though early on Heaviside and Kennelly addressed the effects of the ionosphere on radio propagation and a number of researchers examined the cavity in the intervening years.~\cite{besser}.  I suggest a fitting solution to attribution would be ``Schumann-FitzGerald resonances.''~\cite{schu-fitz}.\\

The name ``Weizs\"{a}cker-Williams method (of virtual quanta)'' is mainly the  fault of the theoretical physics community.  Certainly, in the mid-1930's  the questions about the failure of QED at high energies were resolved by the work of Weizs\"{a}cker and Williams, and Williams's Danish Academy paper showed the wide applicability of the method of virtual quanta together with special relativity.  
But Fermi was the first to publish the idea of the equivalence of the Fourier spectrum of the fields of a swiftly moving charged particle to a spectrum of photons in their actions on a struck system. Williams knew that and so stated. The argument will be made that the choices of appropriate reference frame and struck system are vital to the Weizs\"{a}cker-Williams method, something Fermi did not discuss, but Fermi deserves his due.\\

The relativistic equation for spin motion in electromagnetic fields is perhaps a narrow topic chiefly of interest to accelerator specialists.  It is striking that it was fully developed by Thomas at the dawn of quantum mechanics and before the discovery of repetitive particle accelerators such as the cyclotron. He was surely before his time. Bargmann, Michel, and Telegdi were of an other era, with high energy physics a big business. The cuteness of the acronym BMT and the prestige of the authors made searches for prior work superfluous.  Although the use of ``BMT equation'' is common enough, it is encouraging that in the accelerator physics community the phrase ``Thomas-BMT equation'' is now frequently used in research papers and in reviews and handbooks.~\cite{montague, chao}\\

The zeroth theorem/Arnol'd's law has some similarities to the "Matthew effect."~\cite{merton}    The Matthew effect describes how a more prominent researcher will reap all the credit even if a lesser known person has done essentially the same work contemporaneously, or how the most senior researcher in a group effort will get all the recognition, even though all the real work was done by graduate students or postdocs. The zeroth theorem might be considered as the first kind of Matthew effect , but with some time delay, although some examples do not fit the prominent/lesser constrain.
Neither do my examples reflect, as far as I know, the possible influence by the senior researcher or friends to discount or ignore the contributions of others. 
The zeroth theorem stands on its own, examples often arising because the first enunciator was before his/her time or because the community was not diligent in searching the prior literature before attaching a name to the discovery or relation or effect.\\

\textbf{ACKNOWLEDGMENTS}\\

This article is the outgrowth of a talk given at the University of Michigan in January 2007 at a symposium honoring Gordon L. Kane on his 70th birthday. I wish to thank Bruno Besser~\cite{besser} for \emph{his} citation of my Ref. 1; it drew my attention to E. P. Fischer's adroit characterization of the phenomenon described here. I also thank Mikhail Plyushchay for pointing out Arnol'd's law and Michael Berry for elaboration. And I thank Anne J. Kox for reminding me of Wiechert's contribution to the story of the Lorentz condition.\\ 
\indent For readers wishing to learn more about the scientific work of the neglected, I suggest for Ludvig Lorentz, an article by Helge Kragh;~\cite{kragh} for Emil Wiechert, an article by Joseph F. Mulligan;~\cite{mulligan} for Oliver Heaviside, the book by Bruce J. Hunt~\cite{hunt}; and for George F. FitzGerald, Hunt's book~\cite{hunt} and FitzGerald's collected works, already cited.~\cite{nature}  For Schumann, Besser's paper~\cite{besser} is the obvious source. For the others, better known, a search in library catalogues or on Google will yield results. \\ 
\indent This work was supported by the Director, Office of Science, High Energy Physics, U.S. Department of Energy under Contract No. DE-AC02-05CH11231.

%Begin bibliography


\begin{thebibliography}{99}
\bibitem{schu-fitz} I put FitzGerald's name second because, after all, he treated only the lowest resonant mode.
\end{thebibliography}
\end{document}